# Cross-Participants : fostering design-use mediation in an Open Source Software community

**Flore Barcellini**

Eiffel Team "Cognition and cooperation in design" INRIA & Ergonomics Lab, Research Center on Work and Development, Cnam

Domaine de Voluceau
78153 Le Chesnay Cedex
flore.barcellini@inria.fr

**Françoise Détienne**

Eiffel Team "Cognition and cooperation in design" INRIA

Domaine de Voluceau
78153 Le Chesnay Cedex
francoise.detienne@inria.fr

**Jean-Marie Burkhardt**

Ergonomics –Behaviour & Interactions Lab, Univ. Paris 5

45 rue des Saints-Péres
75006 Paris
jean-marie.burkhardt@inria.fr

**ABSTRACT**

**Motivation** – This research aims at investigating emerging roles and forms of participation fostering design-use mediation during the Open Source Software design process

**Research approach** – We compare online interactions for a successful "pushed-by-users" design process with unsuccessful previous proposals. The methodology developed, articulate structural analyses of the discussions (organization of discussions, participation) to actions to the code and documentation made by participants to the project. We focus on the user-oriented and the developer-oriented mailing-lists of the Python project.

**Findings/Design** – We find that key-participants, the cross-participants, foster the design process and act as boundary spanners between the users and the developers' communities.

**Research limitations/Implications** –These findings can be reinforced developing software to automate the structural analysis of discussions and actions to the code and documentation. Further analyses, supported by these tools, will be necessary to generalise our results.

**Originality/Value** – The analysis of participation among the three interaction spaces of OSS design (discussion, documentation and implementation) is the main originality of this work compared to other OSS research that mainly analyse one or two spaces.

**Take away message** – Beside the idealistic picture that users may intervene freely in the process, OSS design is boost and framed by some key-participants and specific rules and there can be barriers to users' participation.

## INTRODUCTION & MOTIVATION

Open Source Software (OSS) is software that can be run, distributed, studied, changed and improved by its users. OSS design process is characterized by a communitarian and highly mediated design process. This new way of designing is becoming more and more important since there are thousands of OSS (famous examples are the Linux operating system or the Firefox internet browser.) and millions of users of them. These users are considered as the main force of OSS design process compared to proprietary ones: most bugs are detected and fixed because "they are many eyeballs looking at the problem" (Raymond, 1999). In this research, we want to investigate forms of participations and roles' emergence in OSS design process. Our specific focus is on emergence of specific roles fostering mediation between users and developers communities.

After highlighting some features of the OSS design process, we present our research question, and our study on key forms of participation to enhance design-use mediation process in OSS communities.

## OPEN SOURCE SOFTWARE DESIGN

### Communities working in a distributed, mediated and asynchronous form of design

Mainly mediated by Internet tools, OSS design is distributed among three spaces: a discussion space (mailing-lists, forums, chats…), a documentation space (project-related documents archived on the net) and an implementation space (different versions of the software source code) (Sack et al., 2006; Mockus et al., 2002). Thus, it is a paradigmatic case of distant and asynchronous collaborative design which has been less investigated, so far, than distant and synchronous, or co-located collaborative design (e.g. Olson & Olson 2000).

OSS projects are seen as online epistemic communities (Preece, 2000; Cohendet et al., 2000). Their members form a group of people connecting together on the Internet with a common goal- to develop software and to produce knowledge about the artefact they develop. Their activities are framed by implicit and explicit rules: volunteer participation, evaluation of work by a peer-review mechanism of works (e.g. Raymond, 1999). The organization of the design process and roles are emerging from interactions rather than prescribed and formalized *a priori*.

### Statuses and participation of users in OSS design

Different statuses are outlined in OSS projects. Some participants can modify directly the source code, they participate directly to the design process and to the decisions: *(1)* the *project leader* (generally the creator of the project as in Python); *(2)* the *core team* or *administrators*, who have to maintain the code base and the documentation; *(3)* the *developers* who participate to the evolution of the OSS and maintain some of its parts.

Others participants are called "users". In an OSS context, users can be highly skilled in Computer Science, being far away of the classical notion of "end-users". They are called *active users* if they participate in mailing-lists discussions as informants for newcomers, by reporting or correcting bugs with patches, and by proposing new modules. Other users are called *passive users* as they only use the software or lurk on the discussions and documentation spaces of the project.

Forms of participation in OSS projects are supposed to be « open » in time – design becomes continuous, new functionalities can always be proposed and discussed whatever the step in the project (Gasser et al., 2003)- and "open" for different kinds of participant whatever their status (administrators, developers, or users). Users can be potentially involved in all phases in the design process (elicitation of needs and requirements, design and implementation), at least if they have the skills to do so.

### DESIGN-USE MEDIATION PROCESS THROUGH BOUNDARY SPANNERS ROLES

Organizational sciences and design studies describe design as a mutual learning process between users and designers (e.g. Béguin, 2003) in which key-participants can act as boundary spanners or mediators between users and developers groups (Sonnenwald, 1996; Bansler and Havn, 2006) or as brokers between communities of practices (Wenger, 1998).

Boundary spanners are literally persons who span the gap between their organization and external ones (Sarant, 2004). The role of boundary spanners is defined as "communication or behaviour between two or more networks or groups" (Sonnenwald, 1996). They move among different teams transferring information about the state of the project.

This role has been studied in different situations, for instance, in research and development (e.g. Sarant, 2004), in expert work activities (Engeström et al., 1995) and in design situations (Grinter, 1999; Krasner et al., 1987). This is not a formal role but rather a role emerging from interactions and practices. Becoming boundary spanners implies to have developed skills and competencies in the different fields that are spanned. Boundary spanners are well aware of all practices and have gained in legitimacy and credibility in the domains they span.

Boundary spanners appear to be of particular importance in collaborative design situations and they tend to produce innovative solutions to design problems, (Sonnenwald, 1996). They are also key participants who can enhance coordination through informal communications (Krasner et al., 1987). They reduce the amount of information lost or miscommunicated between different phases of design development and different development teams. In non-design situations, it has been described that they can also protect their organisation against external pressure or can represent their organization (Sarant, 2004) and that boundary spanning is a way to develop a horizontal expertise and collective concept formation (Engeström et al., 1995)

As far as we know, in OSS design, and more generally in mediated, asynchronous and distant design situations, the role of boundary spanners has not been investigated yet. Our hypothesis is that design-use mediation may be supported by boundary spanners roles emerging from interactions between users and developers communities of the project. We plan to investigate this issue with a cognitive ergonomics approach, by a specific focus on participants activity, reflecting their "effective role" – distinct from their formal roles or statuses- in the design process (Baker et al., 2003). Although the effective role is dependent upon the formal position of the participant in the community, i.e. the power he/she has on the artefact being designed, it is also contingent upon the participant's actions in the design process (Mahendran, 2002; Sonnenwald, 1996).

### RESEARCH QUESTION AND STRATEGY

#### Forms of participation in OSS design

Our research question targets how use and design are articulated when new functionalities are proposed, solutions are generated and evaluated; and what are the links between users and developers in OSS communities. Our hypothesis is that key-participants can act as boundary spanners between users and developers subcommunities enhancing this way the design-use mediation process.

Indeed, the literature on OSS identify clearly, on one hand, the role of active users participating in the evaluation phase of design (bug reporting and patching, e.g. Ripoche and Sansonnet, 2006) and, on the other hand, the role of the project leader, administrators and developers in the proper design process, that is to say their participations in generating, evaluating solutions and in taking decisions (Barcellini et al., 2005). Open issues are still to characterize participation of project leader, administrators, developers and active users during the whole design process, i.e. from elicitation of needs and requirements phase to the proper design process. Articulating design and use in this distant and asynchronous kind of design can be of particular interest assuming the lack of usability issues in OSS design (Twidale and Nichols, 2005).

#### Research strategy

*Python community and interactions spaces*
This research is focused on a major OSS project called Python, which is a programming language. This project has a large community of users in various application domains and a stable core group of developers who are designing the *Python Core language* and its *standard library* (Figure 1.). One goal of this research is to understand how needs from application domains may impact and enhance the design of the Python core language (symbolized by arrows in Figure 1.).

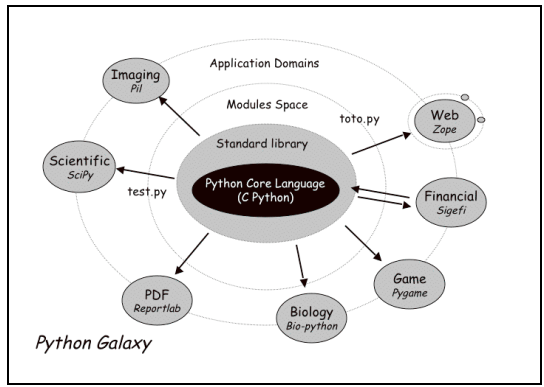

**Figure 1**. The Python galaxy

In a previous paper (Sack et al., 2006), we assume that OSS design is distributed among three online interaction spaces: the *discussion*, *documentation* and *implementation* spaces. In this research, we want to investigate participation in these three spaces articulating discursive, social and technical forms of participation.

The major part of OSS design occurs in the *discussion* space composed of different mailing-lists or newsgroups (Barcellini et al., 2005; Mockus et al., 2002). As all discussions are publicly available and archived online, they constitute rich traces of the OSS design process. In the Python project there are two main mailing lists:

- A *user-oriented mailing-list*, the *python-list* mailing-list is about general discussions and questions about Python. Posting to this mailing-list is the main mean for an active user to participate to the Python project. Most discussions on python-list is about developing with Python, not about development of the Python language itself.

- A *developer-oriented mailing-list,* the *python-dev* mailing-list is for work on developing Python (fixing bugs and adding new features to the Python language). This is the heart of Python's development. Anyone can subscribe to python-dev, though his/her subscription will have to been approved. The list address accepts e-mail from non-members, and the python-dev archives are public.

The *documentation space* is composed by all documents, website, wikis relative to the project. For instance, the Python community has set up a process to propose and document the introduction of new feature into the language the *Python Enhancement Proposal* (PEP). Thus, the *documentation* space contains all the PEP document and their versions.

Finally, the *implementation space* contains the proper code of Python: the standard library and its modules their versions and the traces of their revisions.

*A study of successful and unsuccessful design proposals in the three interaction spaces*

The work presented here aims at comparing a successful "pushed-by-users" design process, investigated in a previous research (Barcellini et al., 2006), to previous unsuccessful attempts to push and resolve the same design problem. These proposals concern the introduction of a decimal type in Python and its related *decimal.py* module.

There have been several unsuccessful proposals, i.e. not scored by an accepted PEP and its implementation, to introduce a decimal type in Python. A first *decimal.py* module was also proposed but was still to be implemented. Then, a successful proposal was initiated by a *user* of Python called the *user-champion*, who is a *developer* of a project in the financial application domainof Python (http://sourceforge.net/projects/sigefi). The first need of the user-champion was for a *money* type, but it appeared that before introducing a *money* type, work had to be done in the *decimal* type in Python. The *user-champion* formalized this proposal through a PEP document that became an accepted PEP (PEP 327). For a more details description of the PEP 327 design process referred to our previous publication (Barcellini et al., 2006).

In the *discussion* space, our objective is to compare the successful and unsuccessful design proposals. We will analyse discussions related to the successful and unsuccessful proposals in both the user-oriented mailing-list (python-list) and the developer-oriented mailing list (python-dev) and compare them according to their temporal organisation and the dimension of participation (*regularity* of the participation, presence of *common participants* and *cross-participants* between the user-oriented and the developer-oriented mailing-lists).

In the *documentation* and *implementation* spaces we will analyse participation via the modifications' actions and who made them. The different versions of the PEP related to the successful design process we study (PEP 327) in the *documentation* space and the different versions of the *decimal.py* module related to the design proposal we study (which is the live implementation of the PEP 327) in the *implementation* space.

Our hypothesis is that *key-participants* can foster the design process, making it successful, and can act as boundary spanners between users and developers enhancing the design-use mediation process. In this direction, we will investigate more deeply identified key-participants roles in the three spaces, in terms of: the social network in which they are involved in the discussion space, and their actions in the *documentation* and the *implementation* space through the revisions they performed.

## METHOD
### Data collection
In the *discussion space*, all discussions are publicly available and archived online (http://mail.python.org/pipermail/python-list/ or python-dev). The data was gathered by searching, by hand, from the python-list (users-oriented mailing-list) and the python-dev mailing-list (developers-oriented mailing-list) for the keywords contained in the subject header of all messages: decimal, money, currency, fixed-point (old solution to resolve decimal problems) and the name of the user-champion. The search was performed from the first formalization to introduce a decimal module in May 2001 to May 2006 (when we began this study). Each message, which was gathered, required reading by the first author to ensure it was indeed a message dealing with the design issue. If the discussion was emerging from another thread we collected this mother-thread.

The data collected are split in two corpora described in Table 1.

- The first one is related to previous unsuccessful attempts called *unsuccessful decimal proposals* corpus (from the first formalization of a decimal module in May 2001 to October 2003, the day before the 1$^{st}$ post of the user-champion).

- The second one is related to the successful design process, called *successfull decimal proposal* corpus (from the 1$^{st}$ post of the user-champion in October 2003 to May 2006).

**Table 1. : Description of the two corpora**

|  | Unsuccessful decimal proposals | | Successful decimal Proposal | |
|---|---|---|---|---|
|  | *Py-list* | *Py-dev* | *Py-list* | *Py-dev* |
| *Nb Discussions* | 10 | 6 | 22 | 29 |
| *Nb Participants* | 66 | 22 | 95 | 48 |
| *Nb Messages* | 192 | 122 | 340 | 406 |

In the *documentation and implementation spaces*, Python code and PEP document revisions (modifications) are also archived and available online (http://svn.python.org/view/). We searched in the various folders for the *decimal.py* module and the PEP 327 modifications. It concerned only the *successful proposal* corpus as: none PEP dealing with decimal were accepted before PEP 327; and because the first version of decimal module, announced in May 2001, was in the "sandbox" of the project, i.e. as to be done.

Concerning the *decimal.py* module; we collected 44 revisions from the 1$^{st}$ of July 2004 (first revision) to the 11$^{th}$ May 2006 (end of the collected data).

Concerning the PEP 327 document, we collected nine revisions from the 29$^{th}$ of January 2004 (acceptance of the PEP ) to the 27$^{th}$ June 2005 (last archived revisions).

## Temporal and design-related organisations of discussions
We characterize the organisations of discussions, according to several dimensions.

The *global organisation* of discussions in parallel in the two mailing-lists and for the two corpora. For each discussion of the two corpora, we identify by hand the first and the last messages and extract the corresponding dates, and the subject-header of the discussions. We then position each discussion along a timeline.

The *design-step organisation* of discussions according to time, i.e. group of discussions dealing with same online design steps: elicitation of needs, proposals, pre-PEP, PEP design, refinements, valorisation of the implemented module, tutorials, debug and evolution. These groups are constituted according their subject-header and a rapid content overview.

The *temporal delay* between discussions' opening within each mailing-list in each corpus. This is an indicator of the follow-up of the proposals' discussions. It corresponds to the means of the delay between the beginning's date of each discussion.

The *presence of parallel same-topic discussions* in the two mailing-lists as an indicator of the thematic coherence between the user-oriented and the developer-oriented mailing-lists. They are same-subjects discussions occurring in the two mailing-lists, that overlap in time. To identify them, we compare subjects and dates of discussions in the two mailing-lists.

## Participations to the three interaction spaces
In the *discussion space*, to highlight the participation we identify different dimensions of participation that can reveal some key-participation.

The *regularity of participation*, for each mailing-list within each corpora. *Regularity* is defined according the number of discussions in which participants are involved in a particular mailing-list. *Regular* participants are those who participate in more than that the third quartile value of the number of discussions in which participants are involved (Q3=2 in the *unsuccessful proposals* corpus, 1 in python-list and 2 in python-dev for the *successful proposal* corpus). Others are called *occasional* participants.

The presence of *common participants* within the two mailing-lists in a same corpus. *Common participants* are those who are present in both python-dev and python-list. To identify these participants we compare between the mailing-lists, the name of posters for each discussion of the corpora.

A sub category of common participants is *cross-participants* between the users and the developers' mailing-lists in the two corpora. We define *cross-participants* (an extended notion of cross-posters, Kollock and Smith, 1996) as persons who participate at parallel same-topic discussions, occurring in the two mailing-lists. To identify the cross-participants, we compare the name of the posters in the same-topic parallel discussions.

The *involvement* of participants in each mailing-list and in each corpus. We define the *involvement* as mean of messages posted by each category of participants (Project Leader, Regular only, Occasional only, Common and Cross-participants).

Finally, we investigate the *social network* in which identified key-participants are involved. This is done on the basis of the quotation patterns between messages posters: i.e. *who is quoting whom*, to highlight interactions between participants. Quotation, i.e. the integration of a part of a previous message(s) in another one is a compensatory linking strategy used by participants to maintain the context in online discussions (Herring, 1999). In a previous paper (Barcellini et al 2005), we have shown that the organisation of messages according to the quoting link is relevant to reconstruct the thematic coherence of online discussions and to understand the interactions between participants in terms of verbal turns. Blocks of quotation are identified in each message by the brackets symbol ">". Then, we search manually in the corpus from which message and which author each quote comes from to obtain a table of "who is quoting who".

In the *documentation and implementation spaces*, we collect the name of the participant who is proposing the revision in the PEP 327 document and the *decimal.py* revisions and explicit references to other participants' works in the content of the revision. Indeed as some rights control the CVS access, developers may modify the code on behalf of other ones. In this situation, the social rules of OSS communities fix to acknowledge other works.

This structural analysis has been complemented by an e-mail interview with the user-champion.

## RESULTS

### Temporal and design-related organisation of discussions

We constructed temporal views of the unsuccessful and the successful design processes (respect. Figure 2 and 3). Each discussion in python-dev and python-list is represented, in parallel, along the timeline by a circle. They are labelled using the subject of the discussions or their corresponding design-step for groups of discussions (black circle).

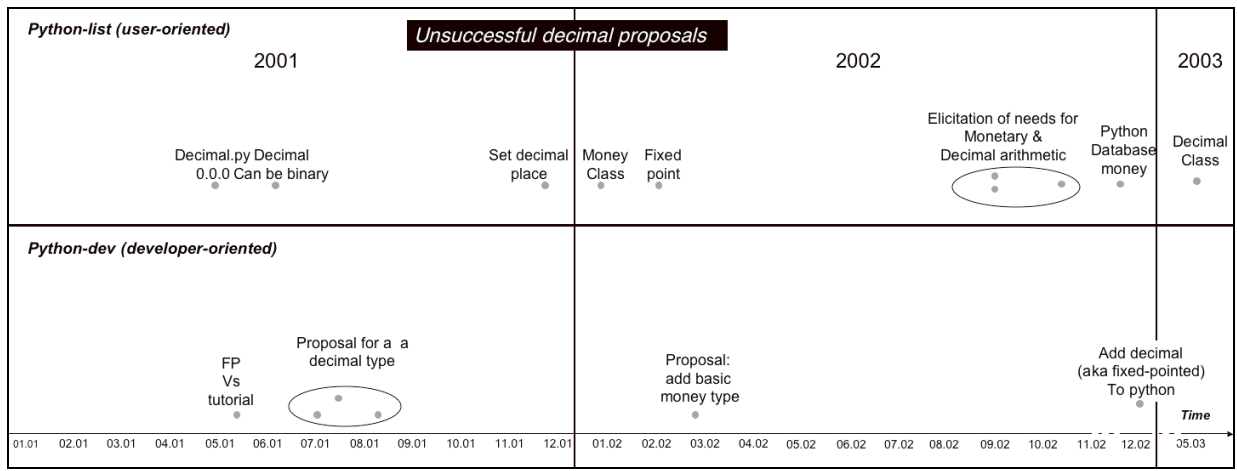

**Figure 2**. Representation of discussions of the *unsuccessful decimal proposals* corpus in python-list and python-dev mailing-lists

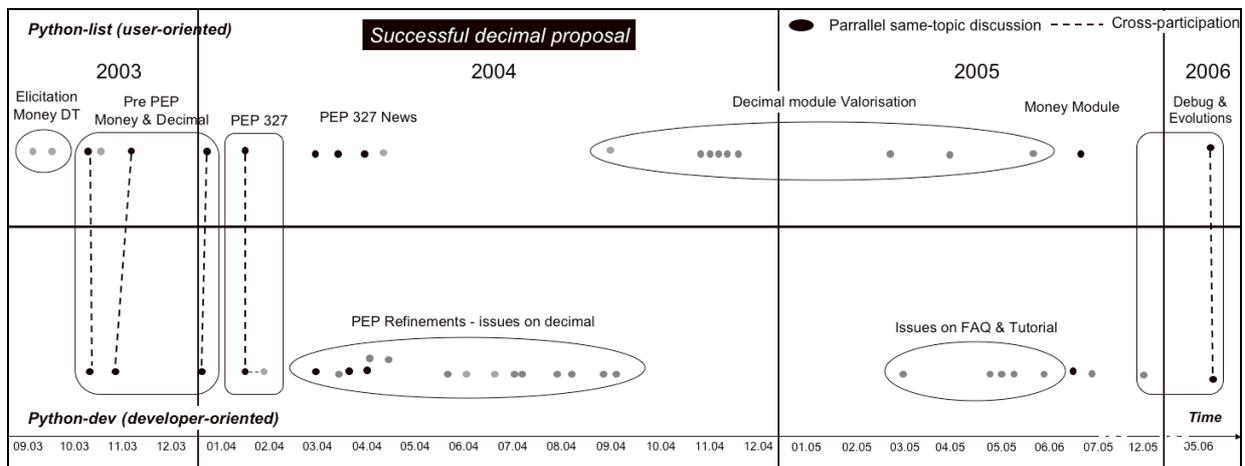

**Figure 3**. Representation of discussions of the *successful decimal proposal* corpus in python-list and python-dev mailing-lists

In the unsuccessful decimal proposals (Figure 2), we identify three attempts to propose a decimal type, in Python-dev (*Pep adding a decimal type to Python* in July 2001; *Proposal: add basic money type* in March 2002 and a *Add decimal (aka fixed point) to Python* in December 2002). They are not followed by same-topic discussions neither in Python-dev nor in python-list. Two design-step-related groups of discussions emerge (*Proposal for a decimal type* and *Elicitation of needs for monetary and decimal arithmetic*) but there are local and are not followed by other steps of a design process (Pre-PEP or PEP) as they are not accepted.

In the successful decimal proposal (Figure 3), there are more structured discussions: the first discussion *Elicitation for a Money DT* is followed, by groups of discussions dealing with the design-steps. Indeed, the mean temporal delay between following discussions is about 30 days in the *successful proposal* corpus and 63 in the *unsuccessful* one. This means that there is more follow-up of discussions in the successful proposal than in the unsuccessful ones.

Moreover, there are five parallel same-topic discussions (labelled using vertical ached lanes) in the successful proposal corpus whereas there are none in the unsuccessful one.

These temporal analyses of the discussions allows us to outline two mains specificities of the successful design process: *(1)* There are parallel same topic discussions in the two mailing-lists; *(2)* we can form design-steps-related group of discussions highlighting the design continuity of discussions in mailing-lists.

**Participation to the discussion space**
*Regularity of participation in the corpora*
Table 2 describes the *regularity of participation* for each mailing-list in each of the two corpora.

**Table 2. : Distribution of participation in each corpus and for each mailing-list**

|  | Unsuccessful proposals | | Successful Proposal | |
|---|---|---|---|---|
|  | *Py-list* | *Py-dev* | *Py-list* | *Py-dev* |
| *Regular Participants* | 18 *(27%)* | 10 *(45%)* | 17 *(18%)* | 14 *(29%)* |
| *Occasional participants* | 48 *(73%)* | 12 *(55%)* | 78 *(82%)* | 34 *(71%)* |
| Total | 66 | 22 | 95 | 48 |

It outlines that they are more participants involved in the *successful* proposal in both the mailing-lits. Most of the participants are *occasional* ones in both mailing-lists and corpora, even if *occasional* participants are less important in python-dev than in python-list. This result highlights a turnover of participants in the mailing-lists, in particular in the python-list.

However, the distribution of *regular* and *occasional participants* is different between the two corpora: the number of *regular participants* is more important in the unsuccessful proposals discussions in the two mailing-lists than in the successful one, especially in python-dev (resp. 45% versus 29%). This result suggests that the presence of regular participants in discussions does not guarantee the design proposal to be successful.

*Common and Cross-participants in the corpora*
In the *unsuccessful proposals* discussions, we identify out of the 84 different participants:

- Four *common participants* (5%): two users, one administrator who is known as a specialist of the decimal domain and who provided the previous solution (fixed-point) to cope with the lack of decimal module in Python, and one developer who proposed the first decimal module (version 0). The two last ones are regular participants in both corpora.

- As there are no same-topic parallel discussions in the *unsuccessful decimal proposals* corpus, there are no *cross-participants* between the two mailing-lists, for this corpus.

In the *successful proposal* corpus, we identify out of the 130 different participants:

- 14 *common participants* (10%): two administrators, five developers, and seven users. Nine are *occasional participants* in both mailing-lists, except four *cross-participants* described below and a user.

- Five *cross-participants* (4%), among the 14 common participants: the user-champion (he was not formally defined as a developer at the beginning of the process and was the project leader of a financial project), an administrator and a developer (also identified as common participants in the unsuccessful corpus), one other developer and one user. All, except the user, are regular participants in the corpus,

In summary, in the *successful proposal* corpus, there are proportionally more *common participants* than in the unsuccessful proposals. Moreover, the successful proposal is characterised by *cross-participants* who appear to be mostly *regular* in their participation.

Table 3 clarifies the *involvement* of participants for each mailing-list and each corpus. It corresponds to the mean number of messages posted by participants in each category.

**Table 3. : Involvement of participants in each corpus and for each mailing-list**

|  | Unsuccessful proposals | | Successful Proposal | |
|---|---|---|---|---|
|  | *Py-list* | *Py-dev* | *Py-list* | *Py-dev* |
| *Project Leader* | 0 | 31 | 0 | 19 |
| *Regular Participants* | 6 | 8 | 7 | 5 |
| *Occasional Participants* | 2 | 1 | 2 | 3 |
| *CC (CrossP)* | 6 *(0)* | 10 *(0)* | 11 *(28)* | 20 *(41)* |

All common participants appear to be more involved in the *successful proposal* corpus, especially in *python-dev* where 20 messages per common participants are posted. The major involvement is from the cross-participants (28 to 41 messages/cross-participants) and the Project Leader but only in *python-dev*. Moreover, the involvement of the Project Leader is stronger in the *unsuccessful proposals* corpus than in the *successful* one. It may outline the different position of the project leader in the unsuccessful vs. successful corpora. In the first case, he was more framing the proposal as he was the unique guarantor of the design process, whereas in the successful proposal this role may be distributed among the cross-participants.

This result highlights that the *cross-participants* seem to be key-participants fostering the design process in the successful attempt, being the most involved and guarantying the follow-up of the design process. Moreover, as two cross-participants were already present (common participant) in the *unsuccessful proposals* corpus, we can argue that the "memory" they have about the design history and rationale may help to foster the design process. In the following, we investigate more deeply the role of these *cross-participants* in the three interactions spaces.

**Roles of the cross-participants in the successful proposal**

*Cross-participants acting as boundary spanners between users and developers in the discussion space*

For the discussion space, the attraction graph in Figure 4 represents who tends to quote whom in both python-list and python-dev, and in which mailing-list. This graph is based on the relative deviation (RD) analysis that measures the association between two nominal variables. It outlines that cross-participants tend to be the link between the users community (U) and the developers community (A-D and PL) with a specific position for the user-champion (U-C, who is also a CP) who quotes and is quoted more by the project leader (PL) and the administrators-developers (A-D) ,i.e. the developers community.

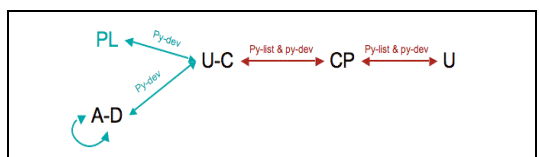

**Figure 4**. Attractions graph of who tends to quote whom

In a previous paper (Barcellini et al., submitted), we outlined, through a message content analysis (activities and references/knowledge sharing) that cross-participants provide knowledge about both the user-oriented application domain and the developer-oriented programming domain (computer-science): this way, they cross the boundaries between users and developers communities acting as *boundary spanners*. The user championing the PEP is a key cross-participant enhancing harmonious social relationships referring to other persons works, and a coordination agent doing synthesis and posting news about the design process in mailing-lists. Moreover, we outlined that he received a social and discursive support as outlined by Figure 4.

*Low contributions of the cross-participants in the implementation and documentation spaces*

In the *documentation space*, *common participants* who are not cross-participants made all the revisions. One unique administrator made five (out of nine) revisions. The four other revisions were made by a specific developer - the PEP editor, who is in charge of the PEP management. In all these four revisions he specified that it was the user-champion contribution.

In the *implementation space*, all the revisions were made by common participants, except one from the project leader: 77% (34/44) of revisions were made by the same administrator than in the *documentation space*, 9% by the user-champion (4/44 from which one effective and three by being referenced), two by the cross-participant specialist of decimals (the specialist of decimal).

However, this low contribution of *cross-participants,* in the implementation and documentation spaces, should be nuanced on the basis of our interview with the user-champion. Indeed, he declared that the three other cross-participants (i.e. except himself and the other user) "helped a lot", such as the administrator strongly present in the implementation space. This technical support could be performed through the embodiment of code in messages or through private exchanges.

**DISCUSSION AND PERSPECTIVES**

Our work provides insights on forms of participation to OSS design process. We outlined that the discussions from the successful design proposal are structured according the design-related-step and between the two mailing-lists. Another result concerns key roles played in this distributed process, the *cross-participants* acting as boundary spanners that relay and support users participation, fostering this way the design-process and guarantying the follow-up of the process.

We plan to extend this work performing interviews with cross-participants and some users involved in this design process to contextualize our data and highlight how their participation is articulated with their activities and needs. We also plan to highlight others conditions, and maybe barriers, i.e. skills needed, to users' participation in OSS communities.

Our contribution is also methodological. Considering the large quantity of data in OSS communities it is tempting, and it is often the case, to use only structural analyses such as social network analysis to characterize OSS design process. We have attempted to illustrate that the combination of structural analyses in the three interactions spaces in which participants are involved is essential to capture the richness and the complexity of the OSSD process. To enrich our findings, this method could be automated (structural analysis) or semi-automated (content analysis).

**ACKNOWLEDGMENTS**

This research is funded, through a Ph. D., by the Cnam, INRIA, and the French Research department.